# Stochastic self-similarity of envelopes of high-frequency teleseismic *P*-waves from large earthquakes suggests fractal pattern for earthquake rupture

*A.A.Gusev*

*Institute of Volcanology and Seismology, Petropavlovsk-Kamchatsky, Russia, and*

*Kamchatka Branch, Geophysical Service,  Petropavlovsk-Kamchatsky, Russia*

**Abstract**

High-frequency (HF) seismic radiation of large earthquakes is approximately represented by *P* wave trains recorded at teleseismic distances. Observed envelopes of such signals look random and intermittent, suggesting non-trivial stochastic structure. Variogram and spectral analyses were applied to instant power calculated from band-filtered observed P-wave signals from eight large ($M_w$=7.6-9.2) earthquakes, with 8-30 records per event and eight non-overvlapping frequency bands analyzed (total frequency range 0.6-6.2 Hz, bandwidth 0.7 Hz). Estimates for both variograms and power spectra look linear in log-log scale, suggesting in most cases self-similar correlation structure of the signal. The range for the individual-event values of the Hurst exponent *H* is 0.71-0.80 (averaged over bands and stations) when estimated from variograms, and 0.78-0.83 when estimated from spectra. No systematic dependence on station or frequency band was noticed. The values of *H* around 0.8 may be characteristic for large earthquakes in general. The result suggests that the space-time organization of earthquake rupture process has significant fractal features. Also, a useful constraint is established for application-oriented earthquake strong motion modeling.

**Introduction**

Teleseismic *P*-waves of large earthquakes, when considered in the high-frequency (HF) frequency range, at frequencies significantly above source corner frequency (typically, above 0.3-0.5 Hz), may bear important information regarding intrinsic properties of propagating earthquake rupture. One simple characteristic of HF fault radiation is its correlation structure. The correlation function of a HF signal itself is not very informative, reminding that of band-limited white noise. Correlation structure of instant HF power, often visualized as wave-train envelopes, is a more promising subject. Short-period teleseismic records of *P*-waves of large earthquakes look intermittent and bursty, and one can hypothesize that they have self-similar, or fractal, structure [Mandelbrot 1983]. This idea is put under test in the following. Simple as it is, it seems to be novel.

Two standard approaches to reveal possible self-similarity of randomly looking observed function are to construct variogram (delta-variance) in time domain, or to estimate power spectral density (PSD) in frequency domain [Mandelbrot 1983]. In case of self-similarity, both these functions must be power laws. Both ways of analysis must produce compatible estimates of the Hurst exponent *H* that is the main parameter that describes a self-similar signal. For variogram, its slope in log-log scale is merely 2*H*. For self-similar, positively correlated stationary signals, PSD behaves as $1/f^\alpha$, where the exponent $\alpha = 2H-1$ is in the range 0-1 (and *H* is in the range *H*=0.5-1). When $\alpha$ is near unity, signals are called flicker-noises ("pink noises"). Our intention is to show that positive flicker noise may be a good representation for instant power signal of seismic wave from a large earthquake.

Attenuation, on the HF side, and rupture duration, on the low-frequency side, define a rather limited frequency range for analyzing self-similarity. Even with relatively long signals from large earthquakes, single-record estimates of *H* and $\alpha$ show significant scatter, and averaging is needed to obtain stable results.

**The technique of data processing**

As initial data the *P*-wave groups were taken recorded by GSN stations at a teleseismic distance. An original velocity record is instrument and (approximately) attenuation corrected. The resulting signal is filtered by a bank of bandpass filters. In the presented version, eight non-overlapping frequency bands are used, jointly covering the 0.6-6.2 Hz frequency range, all with the same -3 dB width of 0.7 Hz; eight central frequencies are from 0.95 to 5.85 Hz. The filter outputs (see Fig 1a as an example) are first used to check the record quality and signal-to-noise (S/N) ratio. If the record is acceptable, the work time window is set interactively.

To reduce non-uniformity related to initial non-random increase and final decrease of the signal amplitude, the beginning of the window is set just before the first clear energetic onset of HF energy (skipping the low-amplitude initial part), and the final point of the window is set so as not to include coda. This preliminary window is then additionally narrowed: its bounds are set between 1% and 97% fractions of the integral of signal power, with integration limits defined by the first window. A common window is used for all usable bands; it is an average version over all of them. Also, a noise window of comparable length is selected before *P* arrival, and noise power is estimated.

After the cutting out, segments of earthquake signal are converted to instant power estimate taken as the squared modulus of analytical signal (SMAS). For all bands, this operation produces smoothed output, with common correlation time of the order of $1/\Delta f \approx 1.4$ s. Noise power is then subtracted from signals, and they are binned (smoothed-and-decimated) with the larger time step *dt* selected to be near to $1/\Delta f$, resulting in nearly-independent samples of instant power. Such decimation is obligatory; otherwise the artificial correlation structure imposed by limited signal bandwidth is imprinted onto the processed trace, and may strongly bias the final result.

Cumulative sums of decimated data were processed according to the standard formula:

$$V(\Delta t) = (1/N) \Sigma (X(t+\Delta t) - X(t))^2 \qquad (1)$$

with $\Delta t = dt, 2dt, 3dt,$ and so on. The resulting empirical variogram was plotted in log-log scale to verify visually the presence (or absence) of a linear trend. Then the value of the slope $2H$ in the relationship

$$V(\Delta t) = C \, \Delta t^{2H} \qquad (2)$$

was determined by linear regression, with weights equal to the numbers of independent samples.

In parallel, frequency domain analysis was performed. We are interested in the value of $\alpha$ in the power-law relationship for power spectral density *PSD(f)*:

$$PSD(f) = C f^{-\alpha} \qquad (3)$$

To determine $\alpha$, one may proceed in a standard way: first estimate *PSD(f)* applying FFT and smoothing; then, perform linear regression on the log-log scale and obtain the estimate of *a*. This approach is known to make relatively noisy estimates however. To stabilize the result for the case of a power-law expected behavior, Pisarenko and Pisarenko [1992] proposed to integrate *PSD(f)* obtaining "cumulative PSD," and then multiply by $1/f$, to recover the original power law exponent. We denote the result *MCPSD(f)*, for "Modified Cumulative PSD":

$$MCPSD(f) = \frac{1}{f} \int PSD(f) df \propto f^{-\alpha} \qquad (4)$$

Using MCPSD one can easily verify visually the presence of the power law behavior, and obtain stable estimates of $\alpha$. This is the approach that shall be used below.

### Data and an example of their analysis

To determine $H$ values of HF instant power, data of several recent large earthquakes (Table 1) were used. *P* wave records of BHZ channel of GSN stations were retrieved from the IRIS DMS data center. Most usable records are from stations on the old continental lithosphere, like BRVK, WRAB, FFC, ARU, or ULN.

On Fig 1a the first step of data analysis is illustrated: band-filtered *P*-wave signals, with their duration reflecting mostly rupture duration. Without any clear final feature, this signal is followed by decaying coda. On Fig 1b, the same data are seen after conversion to SMAS, smoothing and decimation. On Fig. 1c one can see the results of data processing in the time domain for the data in the window of Fig 1b. For each band, variograms are shown in the log-log scale, vertically shifted for visual clarity. No systematic deviation from a linear trend is seen. Variograms are fitted by grey lines whose slopes deliver the estimate of 2*H*. On Fig 1d, frequency domain analysis is illustrated. Traditional PSD estimates are shown as thin lines: dashes are for original spectral points, and solid lines show smoothed spectrum. These estimates are evidently noisy. Bold lines show nearly-linear empirical MCPSD functions described above; they produce stable estimates of $\alpha$.

In a significant proportion of examined records, the data quality was not as good as seen in Fig 1a, and visual check-up of data was essential. Simpler cases are those of significant microseismic noise, as on Fig 2a. We did not use records where less than four bands produced traces with an acceptable S/N ratio. Rejection of noisy traces was performed automatically. In addition, a significant proportion of records is completely unreliable, at least in the high-frequency range (Fig 2bc). In these two and many other cases, spurious HF signals seem to be provoked by the amplitudes of the large event. In some cases, *T* phase superposes on *P* group making traces unusable (Fig. 2d).

For the 2003.09.25 Hokkaido event shown in Fig. 1, the HF source function consists of a single big burst followed by a long tail (of source origin, not scattered coda). One may suspect that this kind of envelope shape may automatically create significant long-range correlation. This was found untrue: power-law correlation structure was revealed for a few events with multiple bursts of energy. The 2004.12.26 Sumatra mega-earthquake is of the same kind, but it created a more complicated pattern.

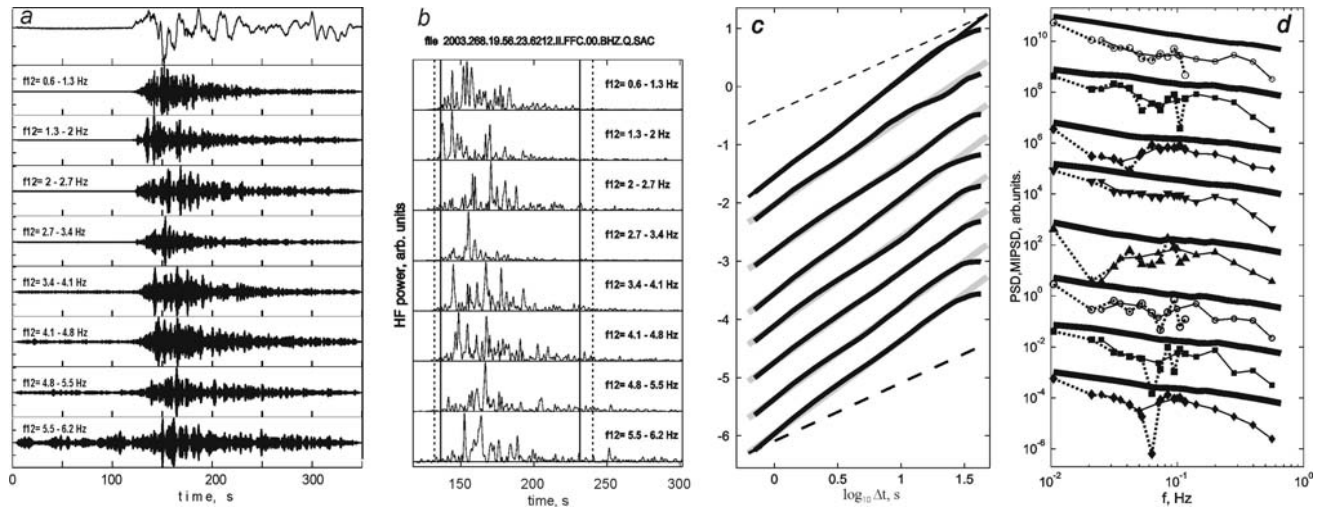

Figure 1. (*a*) Bandpass-filtered traces of *P*-waves of the 2003.09.25 *M*=8.1 Hokkaido earthquake at BHZ channel of FFC at $\Delta$=68°; the uppermost trace is broadband displacement, plotted for general orientation; band central frequencies are labeled. (*b*) Binned and decimated SMAS (instant power) for the same bands. Vertical dashes indicate bounds for the preliminary work window, and solid ones delimit the final one. (*c*) Variograms of windowed traces of *b*, bands in the same order from top to bottom. Grey lines: linear fits to data, with slopes that deliver the estimates of 2*H*. For orientation, dashed lines are given with slope 1.0, expected for uncorrelated data. (*d*) Spectral analysis of same traces. Symbols and thin lines represent regular PSD estimates, see text for details. Bold lines are preferred MIPSD estimates.

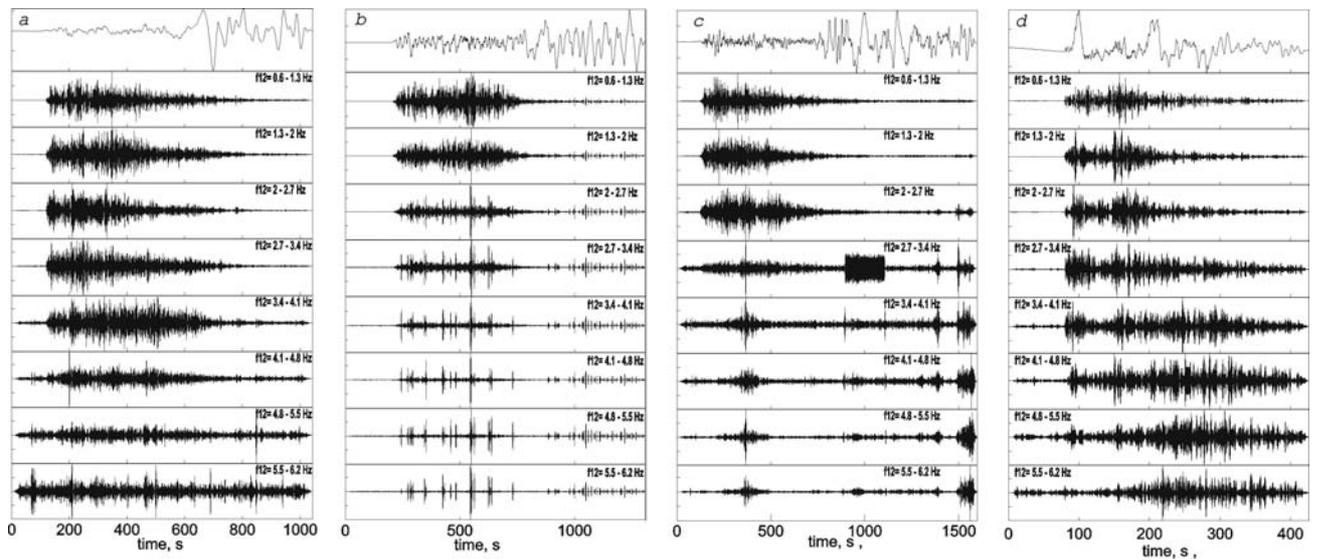

Figure 2. Examples of rejected data. (*a*) Event 2004.12.26 at KMBO, three lowermost traces are rejected automatically, the rest ones are qualified as usable. (*b*) Same event at KIV, spurious pulses of presumably instrument/digitizer origin. (*c*) Same event at SBA, spurious signals of unclear origin. (*d*) Event 1998.03.05 at WRAB. Powerful *T*-phase signal is seen above 3Hz.

On Fig 3, analogs of Fig 1cde are shown for this earthquake as recorded at SNZO, and also spectral plots for KONO, ALE, and ARU. One can see that self-similar signal structure is clearly seen over the $\Delta t$ range of 1.5-100 s, as well as in the corresponding frequency band 0.01-0.6 Hz, but there is more or less clear saturation at longer delays/lower frequencies.

The possibility was checked that the power-law behavior is caused by propagation effects. That is, the combination of *P* group with *P* coda might produce the power-law spectral/variogram structure alone. To examine such a possibility, a number of records of smaller-magnitude events were analyzed, but no spectral imprint of self-similarity was seen. This is illustrated on Fig 4abc with analogs of Fig 1abc for the $M_w$6.8 Peru event of 1998.12.14 recorded at WRAB. To illustrate performance of the entire processing procedure, spectra are also shown for two synthetic input signals. For the case of stationary white noise in Fig 4d, MCPSD spectra show no decay with frequency. In Fig 4e, similar spectra are given for the case of white noise modulated by deterministic envelope shaped as an isosceles triangle. Now MCPSD spectra decay with increasing frequency, but spectral trends are evidently concave in log-log scale, not linear.

**Average and individual estimates of *H* and $\alpha$, and components of their dispersion**

The processing procedure described above was applied to records of eight large earthquakes (Table 1). Five events are from subduction zones, the rest are crustal events. Eight to 32 records were processed for each event; for higher-frequency bands the number of usable traces was lower. Estimates of *H* and $\alpha$ were calculated for each record and frequency band; their averages are given in Table 1. Along with *H* estimate proper, the derived value $H(\alpha) \equiv 0.5(\alpha+1)$ is also given. Event averages are accompanied by estimates of inter-record rms deviation $\sigma_{ir}$ for each band and event, averaged over bands. In the following text, for compactness, each estimate related to *H* proper is accompanied by the corresponding estimate for $H(\alpha)$, in parentheses. Average of $\sigma_{ir}(H)$ over bands and events is 0.063 (0.052).

When *H* estimates for records of a certain event is a certain band are averaged, there is dispersion between such averages (inter-band dispersion). It is characterized by empirical rms deviation $\sigma_{ib}$ =0.016 (0.017) (average over events). Analysis of variance shows that this value is insignificantly different from zero, thus the empirical inter-band dispersion is completely explainable by random perturbations caused by inter-record variations. Also, no systematic change of *H* with frequency was noticed (either for individual events, or for the averages). Inter-event dispersion of *H* estimates is highly significant statistically both for *H* proper and $H(\alpha)$, but is not identical in these two cases: rms deviation $\sigma_{ie}$ equals 0.031 for *H* proper, against as low as 0.014 for $H(\alpha)$. The difference of scatter of *H* estimates of two kinds is in fact evident if one looks at the corresponding columns of Table 1.

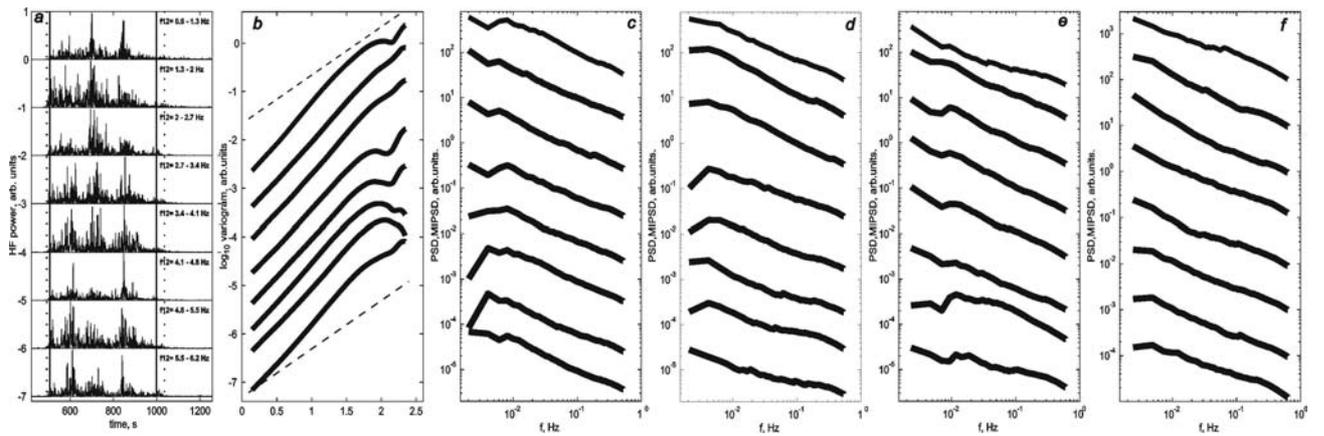

Figure 3. Limited range of self similarity for the 2004.12.26 mega-earthquake. (*a, b, c*)– analogs of Fig 1*bcd* for SNZO, (*d , e, f*) – spectral plots only, for KONO, ALE, and ARU

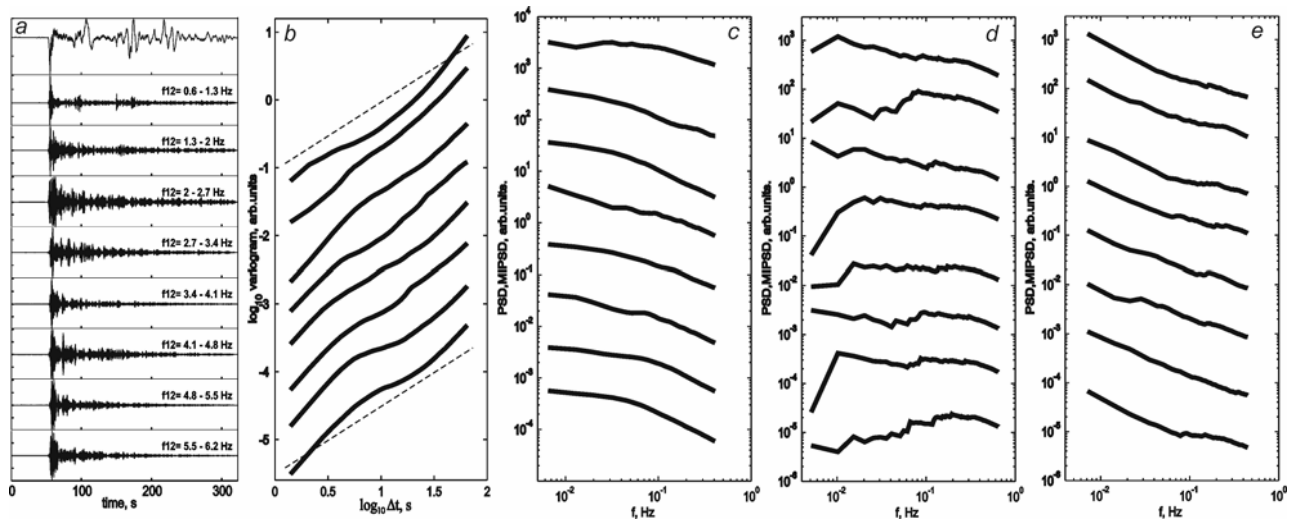

Figure 4. Examples of non-self-similar data. (*a*) bandpass-filtered traces of 150 s of the record after the first arrival of 1998.12.14 event (*M*=6.8) at WRAB. (*b, c*) corresponding variograms and *MIPSD*(*f*). (*d, e*) *MIPSD*(*f*) for 150 s of simulated stationary white noise (*d*) and for similar signal modulated by deterministic triangular envelope (*e*).

8The data regarding variance are rather informative. One can see that inter-record variance $\sigma_{ir}^2 \approx 0.06^2$ is the dominating component that defines dispersion of estimates between various records and bands. Inter-event variance $\sigma_{ie}^2 \approx (0.015-0.03)^2$ is much lower, but real, indicating that average $H$ values do vary among events. One can also note that inter-event rms deviation of 0.015-0.03 amounts to only about 3-6% of the complete range of possible $H$ values for the case of positive correlation, i.e. 0.5 to 1.0, and infer that despite the presence of mentioned inter-event variations, the $H$ value of instant power signal from an earthquake is relatively stable. It shows only a limited variation among different events, and no discernible variation among frequency bands. A finer analysis suggests that inter-record variance is of purely statistical origin, and there are no meaningful differences between true $H$ values for different stations/rays, or bands.

Comparing averages of $H$ and $H(\alpha)$ one can note that there is slight mismatch between these. Its probable cause is the imperfectness of procedures of estimating $H$ in time and frequency domains, because a mismatch of a similar kind and magnitude was found when the described processing procedure was applied to synthetic data sets.

Table 1. Estimates of $H$ and $\alpha$ for eight earthquakes, averaged over records and over frequency bands

| Date | $M_w$ | Region | Type[1] | $N/N_{min}$[2] | $H$ | $\sigma_{ir}(H)$ | $H(\alpha)$ | $\sigma_{ir}(H(\alpha))$ | $\alpha$ | $\sigma_{ir}(\alpha)$ |
|---|---|---|---|---|---|---|---|---|---|---|
| 19971205 | 7.9 | Kamchatka | Su | 19/7 | 0.75 | 0.07 | 0.82 | 0.05 | 0.63 | 0.09 |
| 19980325 | 8.2 | S.Pacific | Cr | 8/2 | 0.79 | 0.03 | 0.83 | 0.04 | 0.66 | 0.08 |
| 20021103 | 8.4 | Alaska | Cr | 17/6 | 0.71 | 0.06 | 0.83 | 0.05 | 0.66 | 0.11 |
| 20030925 | 8.2 | Hokkaido | Su | 18/10 | 0.78 | 0.05 | 0.82 | 0.05 | 0.63 | 0.09 |
| 20041226 | 9.2 | Sumatra | Su | 32/21 | 0.75 | 0.07 | 0.80 | 0.05 | 0.60 | 0.09 |
| 20050328 | 8.7 | Sumatra | Su | 20/11 | 0.80 | 0.04 | 0.82 | 0.04 | 0.64 | 0.10 |
| 20060420 | 7.6 | NE Russia | Cr | 28/4 | 0.72 | 0.08 | 0.78 | 0.07 | 0.56 | 0.14 |
| 20061115 | 8.1 | Kurile Isles | Su | 14/7 | 0.74 | 0.05 | 0.80 | 0.05 | 0.60 | 0.11 |
| *average* | | | | | 0.76 | 0.063 | 0.81 | 0.053 | 0.62 | 0.010 |

[1]Tectonic style of a fault: subduction(Su) or crustal (Cr)

[2]Total number of records used and number of usable records for higher-frequency bands

**Discussion**

It is common to treat the propagating earthquake rupture in frames of the concept of a propagating shear crack, using such notions as fault-average stress drop, cohesion zone width, etc. In such a treatment, only two characteristic times/frequencies typically appear, one related to crack size and another to cohesion zone width. One more characteristic time is rise time, or slip pulse duration [Heaton 1990], but it is somewhat alien to the crack concept. Multiscaled dynamics of a real rupture manifests itself most clearly in the now classical omega-square source spectrum models of Aki [1967] and Brune [1970] that predict power law, i.e. self-similar spectral shapes at frequencies significantly above the corner frequency. However, this model gives only a general idea; real spectra significantly deviate from this model and show non-scaling spectral shapes with humps [Gusev 1983]. A purely kinematical stochastic self-similar space-time fault model was proposed by Andrews [1981]. However, with respect to the time structure of HF record envelopes, this model seems to be oversimplified, as it predicts no self-similar envelope structure (there is only single characteristic time in this model, related to full rupture duration).

Generally, more detailed analysis of temporal and spatio-temporal organization of HF energy can improve our understanding of source processes. For example, recently it was found that correlation of high-slip and high-HF-power-generating areas within the source/fault is only limited [Gusev et al. 2004], against the predictions of most common broad-band earthquake fault models that consists of multiple smaller cracks. Flicker noise behavior of band-passed wave power revealed above gives a direct support to the idea that there is a multiplicity of temporal scales in the earthquake fault process. Note that there is no need to relate each scale to a population of subevents of a specific size and/or duration, as was hypothesized by Blandford [1975].

The obtained results provide an important constraint for development of realistic broad-band models of earthquake rupture formation. In addition, they provide a useful check for validation of simulation codes aimed at realistic imitation of strong ground motion: a reliable model must generate far-field HF radiation with realistic spectra of instant power.

**Acknowledgements**

The study was supported by the Russian Foundation for Basic Research (Grant No 07-05-00775).